# Improvement in the performance of multilayer insulation technique and impact in the rare physics search experiments

D. Singh[1*], M. K. Singh[1,2†], A. Chaubey[1], A. K. Ganguly[3], V. Singh[1,4‡]

[1]Department of Physics, Institute of Science, Banaras Hindu University, Varanasi 221005, India
[2]Institute of Physics, Academia Sinica, Taipei 11529
[3]Department of Physics, MMV, Banaras Hindu University, Varanasi 221005, India
[4]Physics Department, School of Physical and Chemical Sciences, Central University of South Bihar, Gaya 824236, India

## Abstract

Providing thermal insulation to systems at very low temperature from surroundings, involves blocking the transport of thermal energy−regular or enhanced, taking place through radiative, conductive and convective processes. For instance, the enhancement of radiative heat transport that takes place by infra−red or far infra−red light at low temperature is due to diffractive propagation. The wavelength of light in this part of the spectrum usually lie in the range of mm to cms. Hence it can get bent across an obstacle while propagating forward. Apart from radiative, the convective and conductive processes also get affected due to appearance of non−linearities in the modes of lattice vibrations and anomalies in material transport due to the appearance of vorticity and turbulence in the intervening media. The Multilayer insulation (MLI) technique has offered a robust thermal protective mechanism to provide proper insulation to the cold walls of the cryostats from the heat of the surroundings (basically the radiation heat load). This work is focussed on the estimation of performance and efficiency of the MLI technique as well as exploration of its versatile applicability. Three different spacer materials such as Dacron, Glass−tissue, and Silk−net with radiation shields are selected for the intervening medium in the present study. This article explores the thermal performance of MLI system by changing the physical parameters (emissivity and residual gas pressure), varying the geometry of the radiation shields (perforation styles of radiation shields) and by analyzing the effect of arrangement of radiation shields on the conduction heat load. The predictions of analytical models−Modified Lockheed equation, Lockheed Martin Flat Plate equation and McIntosh's approach are compared for the performance of MLI systems and for the applicable choice in future experiments. This analysis is concluded by studying the possibility of using MLI technique in the health sector by reducing the evaporation rate of liquid Oxygen ($LO_2$) during pandemic situations e.g., in COVID−19.

**Keywords**: Multilayer Insulation, Thermal protection system, Temperature gradient, Aerodynamic heating
**PACS Nos.:** 07.20.Mc; 05.70.-a; 07.20.-n; 07.90.+c; 44.10.+i; 44.40.+a; 65.40.Gr

[*]Email:damini.singh13@bhu.ac.in
[†]Email:manu@gate.sinica.edu.tw
[‡]Email:venktesh@bhu.ac.in

# 1   Introduction

Multilayer insulation (MLI) technique is being extensively used in both space and ground based cryogenic programs as a passive thermal protection system because of it's small thermal conductivity, great implication and immediate impact on the heat transfer [1]. The salient feature of MLI technique is it's capability of reducing heat load, in particular, the radiation heat load to the cold walls of the cryostats. This technique is based on the principle of achieving the multiple reflection of incident radiation by placing radiation shields in between the two walls of the cryostat [2]. These radiation shields are also known as reflecting shields because they are made up of highly reflecting metallic materials. Therefore, in order to reduce the conduction between these adjacent radiation shields due to their direct contact, low thermal conductivity (insulators) materials are dispersed in contact with these radiation shields. These low thermal conductivity separator materials are called as spacers having high heat capacity and high service temperature [3]. It follows that the main components of MLI technique are radiation shields and spacers. It is desired to take care of the designing of MLI blanket such that it can minimize the heat load in every manner including edges, seams, and installation procedure. Although these spacers prevent the direct contact between the radiation shields, they themselves are in contact with the radiation shields. This leads to the reduction in thermal conduction between adjacent radiation shields, however, a significant amount of thermal conduction through the first spacers can takes place. Although the incident radiation reflected from the first radiation shield, a small amount of incident radiation moves through the first spacer. Thus, the radiation and thermal conduction both heat load flux moves through the first spacer as well as to the second radiation shield. This process is successively continued till the cold wall boundary and results in a great decrement in the radiation heat load [4]. A schematic diagram representing the working principle of MLI technique is shown in Fig. 1.

Radiation shields are usually made up of thin (paper−like) polyethylene or Mylar sheet reflective layers and permit a larger number of such layers/cm in between the cold and hot wall boundary. These reflective layers are usually coated with Aluminium (also Gold or Silver) on both the sides to achieve the high value of reflective performance [5]. The radiation shields which are typically used in many MLI performances are Double Aluminized Kapton [6], Gold coated ceramic foils [1], Double Aluminized Mylar (DAM) [4] and Aluminium foils [5]. There are many spacers used in various MLI performances like Vegetable fiber, Nomex [6], Carbon fibrous, Alumina fibrous, Amorphous carbon [1], Terylene film, Polyester spacers [2], Dacron, Glass−tissue, Silk−net [7], Refrasil, Dexi glass, Silica fiber felt [5] etc. Radiation shields are coated with Aluminium because of their low cost and high melting point [8].

Practical applications of MLI technique as a thermal insulator have been found in the technology−demonstration missions and the architecture of cryogenic spacecraft [9]. It is being utilize in space exploration programs like spacecraft system designing (NASA's programs) [10] and has been used in several space−science missions such as IRAS [9], COBE [9] launching and in the launching from the ISO organization [9] etc. It is extensively being used in accelerator physics experiments like LHC [11]. Moreover, it has also been applied in high temperature programs in order to protect the sensitive devices from damages due to extreme temperatures like transatmospheric space vehicles [3], hypersonic vehicles to reduce the aero-dynamic heating [12], etc. Cryogenic propellant and high temperature fuel are also using MLI technique for storage purpose [1].

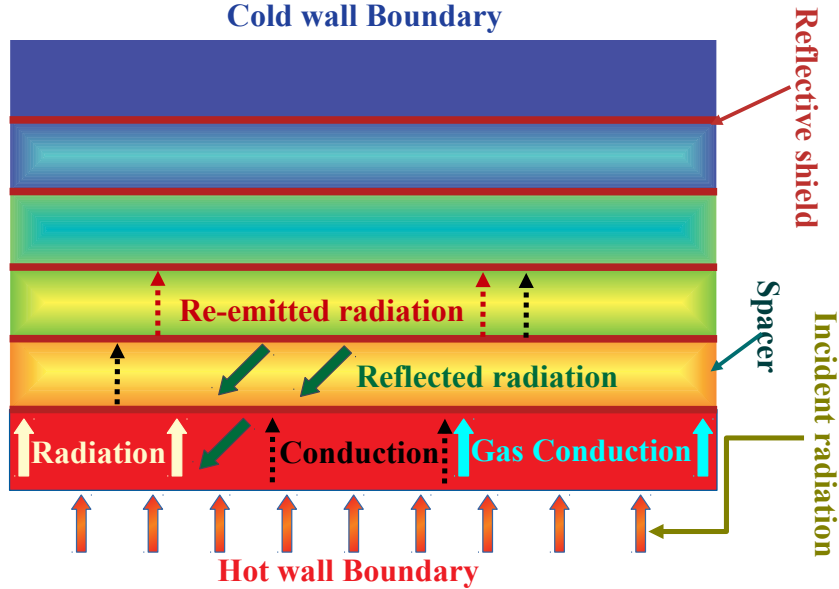

Figure 1: The schematic diagram representing the basic principle of MLI technique, multiple radiation reflection through radiation shields.

The current work focuses on low−temperature applications of MLI technique in current running or upcoming projected tonne−scale fundamental physics research programs, in particular, the dark matter (DM) and neutrinoless double beta decay ($0\nu\beta\beta$−decay) search experiments. MLI technique is being used in the DM search experiments like, EDELWEISS [13], CRESST [14], EURECA [15] and in the search of neutrinoless double beta decay experiments like, KamLAND−Zen [16], nEXO [17], CUORE [18], CDEX [19], GERDA [20] and in future LEGEND which is the union of GERDA and MAJORANA [21].

Heat load reduction at the cold wall of the cryostat leads the cost effective application of MLI technique in each long−time running experiments. Cost effectiveness can be accounted in terms of data taking period using the same amount of cryogenic liquid. There are numerous important physical parameters associated with the MLI technique which might affects its role of heat load reduction. These parameters are emissivity of the radiation shields, residual gas pressure and perforation styles of the radiation shields. Position of the first radiation shield and the arrangement of radiation shields also affects the working of MLI technique. Uncertainty in the evaluation of the performance of MLI systems through the different theoretical models should also be accounted. The present work explores improvement methodologies in the thermal performance of MLI systems. A detailed analysis is performed for the suitable material combination of radiation shield and spacer.

Other than the application mentioned above, this technology can be used in the health sector too. In the recent outbreak of COVID−19 infections that the whole world is facing, approximately 40 % of COVID−19 patients are facing mild illness and about 15 % of them have severe illness requiring oxygen therapy. This leads to the grave shortages of medicinal oxygen in the pandemic. Therefore, in order to fulfil the demands of the same, it highly required to preserve the liquid oxygen (LO$_2$) (usually kept in cylindrical cryostat) from the environmental thermal fluctuations in addition with boost up their commercial production at the same time. The present work also elaborate the application of MLI technique in reducing the evaporation rate of LO$_2$.

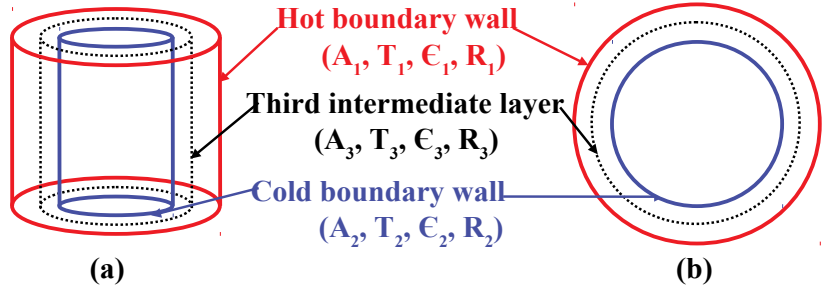

Figure 2: Schematic diagrams of cylindrical (a) and spherical (b) cryostat designs with the insertion of a third intermediate layer between the hot and cold wall boundaries.

## 2 Heat exchange between hot and cold wall boundaries of a cryostat

The impact of MLI technique in reducing the heat load can be estimated by calculating the total heat load (q$^{total}$) to the inner wall of the cryostat. The role of MLI technique in accomplishing it's aim, can be understood by making a comparison of heat loads with and without MLI technique. The heat load without inserting any layer and after inserting a third layer between the two (hot and cold) walls of a cryostat can be evaluated. An elementary introduction to heat transmission and radiative heat load transfer by reflectors are provided in the appendix 6.
In order to understand the basic heat load evaluation methodology, a schematic layout of spherical as well as cylindrical geometries of a cryostat are shown in Fig. 2. Present work considered that the outer and inner walls of the cryostat are buildup with mechanically polished copper [22] and the exchange of heat load is from the outer hot wall to the inner cold wall of the cryostat. Copper is selected for this purpose because of having a low value of emissivity as compared to Aluminium after proper mechanical polishing and it rarely brittles even at very low temperatures [4]. The expression for the heat load q$^{1-2}$ to the inner wall of the cylindrical cryostat (without inserting any layer) can be expressed as [23]:

$$q^{1-2} = \sigma A_2 (T_1^4 - T_2^4) \left[ \frac{1}{\varepsilon_2} + \frac{1-\varepsilon_1}{\varepsilon_1} \left( \frac{A_2}{A_1} \right) \right]^{-1} \quad (1)$$

where, $\sigma = 5.675 \times 10^{-8}$ Wm$^{-2}$K$^{-4}$ is the Stefan−Boltzmann's constant, $\varepsilon_1$ and $\varepsilon_2$ are the emissivities, $T_1$ and $T_2$ are the temperatures (in K), $A_1$ and $A_2$ are the surface areas ($A_1 > A_2$) of the outer hot wall and inner cold wall boundaries, respectively. Typically, the heat trans-fer via radiation between the two enclosed surfaces (such that $T_1 > T_2$) takes place from out−to−inside [23].

The heat load $q^{1-2}$ after inserting a single intermediate layer of the same material with surface area $A_3$ (such that $A_1 > A_3 > A_2$), emissivity $\varepsilon_3$ and temperature $T_3$ (the average temperature of the hot and cold boundaries), in between the two walls of a cylindrical cryostat can be expressed as [4]:

$$q^{1-2} = \sigma A_2 (T_1^4 - T_2^4) \left[ \frac{1}{\varepsilon_2} + \left( \frac{1}{\varepsilon_1} - 1 \right) \frac{A_2}{A_1} + 2 \left( \frac{1}{\varepsilon_3} - 1 \right) \frac{A_2}{A_3} + \frac{A_2}{A_3} \right]^{-1} \quad (2)$$

The heat exchange between the two walls of a concentric spherical cryostat such that $A_1 > A_2$ and $T_1$ and $T_2$ follows the similar expression for heat transfer as in Eq. 1 and Eq. 2 [24].

A minimum amount of the heat load to the inner wall of a cryostat would exhibits a best performing MLI system. There are different theoretical models available to estimate the heat load in different situations. The current work would explore the three such most versatile empirical models to estimate the best performance of MLI technique.

## 2.1 Modified Lockheed Equation

The first analytical approach to investigate the performance of MLI technique is the Modified Lockheed equation which accounted for all the three modes (radiation, solid conduction as well as gaseous conduction) of heat exchange. This is given by [25]:

$$q^{\text{total}} = q^{\text{radiation}} + q^{\text{solid conduction}} + q^{\text{gas conduction}} \quad (3)$$

The expression for the total heat load in the Modified Lockheed equation can be expressed in an empirical form as [26]:

$$q^{\text{total}} = \frac{1}{l} C_r \ \varepsilon \ (T_1^{4.67} - T_2^{4.67}) + \frac{1}{2\,(l+1)} C_s \ \bar{l}^{2.63} (T_1^2 - T_2^2) + \frac{1}{l} C_g \ P \ (T_1^{0.52} - T_2^{0.52}) \quad (4)$$

where $\bar{l}$ stands for layer density and $l$ represents the number of layers. The terms $C_r$, $C_s$ and $C_g$ are the empirical constants known as coefficients of radiation, solid conduction and gas conduction, respectively. Here $C_r$ is a function of reflector's material, $C_s$ is a function of spacer's material and $C_g$ is a function of radiation gas pressure between the radiation shields, respectively, as explained in the appendix 6. Symbol $\varepsilon$ represents the emissivity of the radiation shields and $P$ symbolizes the residual gas pressure [26].

It should be noted that, the "solid conduction" term in Eq.4 would be modified for the Dacron spacer material. This is because in the original Lockheed equation the spacer material used was Glass−tissue with different sizes of radiation shield's perforation, whereas in the calculation, Dacron material was used as a spacer with different sizes of radiation shield's perforation. After incorporating the modification, the Modified Lockheed equation for the total heat load for the Dacron can be expressed as [25]:

$$q^{\text{total}} = \frac{1}{l} \left[ C_r \ \varepsilon \ (T_1^{4.67} - T_2^{4.67}) + c \ \bar{l}^{2.63} (T_1 - T_2) + C_g \ P \ (T_1^{0.52} - T_2^{0.52}) \right] \quad (5)$$

where c = $2.4 \times 10^{-4}(a+b)$ in which a = $(0.017 + 7 \times 10^{-6}(800-T))$, b = $0.0228\ ln(T)$ and $T$ is the average temperature of the hot and cold boundary temperatures. The value of all constants are mentioned in Table 1.

## 2.2 Lockheed Martian flat plate equation

The second analytical approach is the Lockheed Martin Flat Plate equation which is developed with flat plate calorimeter for Silk−net spacers for liquid nitrogen (LN$_2$) and assumes negligible gas conduction under vacuum levels that is less than $10^{-5}$ torr. Thus it includes only conduction and radiation heat loads and the empirical form of the total heat load in the Lockheed Martin Flat Plate equation can be expressed as [27, 28]:

$$q^{\text{total}} = \frac{1}{l} C_r\ \varepsilon\ (T_1^{4.67} - T_2^{4.67}) + \frac{1}{2(l+1)} C_s\ \bar{l}^{2.56}(T_1^2 - T_2^2) \tag{6}$$

where all constants follows their usual meaning as explained in Eq.4 [27, 28] and the value of all constants are mentioned in Table1.

## 2.3 McIntosh's approach

The third analytical Layer−by−Layer approach is the physics−based expression developed by McIntosh for the theoretical calculation of the heat load in MLI system [2, 7, 29]. The heat radiation exchange term in this approach is [2, 7, 29]:

$$q^{\text{radiation}} = \sigma(T_1^4 - T_2^4)\left[\frac{1}{\varepsilon_1} + \frac{1}{\varepsilon_2} - 1\right]^{-1} \tag{7}$$

where all the parameters have their usual meaning as described above and the values of $\varepsilon_1$ and $\varepsilon_2$ is mentioned in Table1. The heat exchange through the gas conduction term is [2, 7, 29]:

$$q^{\text{gas conduction}} = D_g P \alpha (T_1 - T_2) \tag{8}$$

where $D_g = \left(\frac{K_g}{P\alpha}\right)$, in which $K_g$ stands for the gas conduction (in Wm$^{-2}$K$^{-1}$), $P$ is the gas pressure (in pascal, 1 pascal = $7.5 \times 10^{-3}$ Torr), $D_g = \left(\frac{\gamma+1}{\gamma-1}\right)\sqrt{\frac{R}{8\pi MT}}$ and its value is 1.1666 for the air [2, 7, 29], $\alpha$ is the accommodation coefficient (0.9 for air [29]), $\gamma = \left(\frac{C_p}{C_v}\right)$, $R$ is the gas constant whose value is taken as 8.314 kJ−mol$^{-1}$K$^{-1}$, $M$ is the molecular weight of gas (in kg−mol$^{-1}$) and $T$ is the temperature of vacuum gauge (usually ∼300 K). The heat exchange via solid conduction can be represented as [2, 7, 29]:

$$q^{\text{solid conduction}} = K_s(T_1 - T_2) \tag{9}$$

where $K_s = \left(\frac{D_s f k}{\Delta x}\right)$, in which $D_s$ is an empirical constant and taken as 0.008 for Dacron [7, 25, 29], $f$ is the relative density of the separator compared to solid material (0.02 for Dacron [29]), $k$ is the separator material conductivity (in Wm$^{-1}$K$^{-1}$) which is considered in the range from 0.15 to 0.27 Wm$^{-1}$K$^{-1}$ [30] and $\Delta x$ is the actual thickness separator between reflectors (in meter). This solid conduction part of McIntosh's approach is fitted for Dacron with reasonable accuracy [7, 25, 29]. It follows that the total heat transfer in McIntosh's approach comes out to be the sum of Eq. 7, Eq. 8 and Eq. 9 and expressed as [2, 7, 29]:

$$q^{\text{total}} = \sigma(T_1^4 - T_2^4)\left(\frac{1}{\varepsilon_1} + \frac{1}{\varepsilon_2} - 1\right)^{-1} + D_s f k (T_1 - T_2)\frac{1}{\Delta x} + D_g P \alpha (T_1 - T_2) \qquad (10)$$

It is obvious from Eq.4 and Eq. 10 that the Modified Lockheed equation and the Layer−by−Layer McIntosh's approach accounts for all the three modes of heat exchange as expressed in the Eq. 3.

Although all these modes of heat exchange follows a variation from layer to layer, the total heat load remains constant throughout the whole MLI blanket. McIntosh's approach follows the flow of heat from layer to layer, which is different from the Modified Lockheed equation which considers the flow of heat over the entire set of MLI blanket [31]. This is because of the presence of $(l+1)^{\text{th}}$ term in the denominator of Eq. 4, which requires bulk information instead of Layer−by−Layer to perform the analysis. It follows that the Modified Lockheed equation should not be used to perform the Layer−by−Layer analysis, whereas it is preferred for the bulk analysis. However, the McIntosh's approach must be follows to analyze the heat load at each layer [26]. If the number of layers is sufficiently large then the term $(l+1) \approx l$ can also be approximated [32, 33].

Selection of appropriate material is quite cumbersome to obtain a best MLI system. This is because the selection criteria for a suitable combination of material not only focuses the thermal properties, it also consider the mechanical properties. In the current work, Perforated DAM with Dacron, Perforated DAM with Glass−tissue and Unperforated DAM with Silk−net are selected for further studies as a suitable material combination of radiation shield and spacer. Empirical coefficients and relevant constants for these materials are given in the Table 1. A variation of the heat load with hot boundary temperature in the above discussed three models is shown in Fig. 3. In order to illustrate this comparative variation, the value $\bar{l}$ is selecetd to be 20 layers/cm [34] for a MLI blanket with thickness of 20 mm [35]. Typically, in most of the experiments the cryostat is surrounded by water tank [20] at room temperature. It follows that the value of $T_1$ is considered as 300 K [36] (water at normal temperature) and the value of $T_2$ is taken as 77.4 K for Liquid Nitrogen (LN$_2$) [37–39].

It is evident from the above discussion that the Lockheed Martin Flat Plate Eq. 6 is valid only for Silk−net spacers [27, 28] and the layer by layer model developed by McIntosh Eq. 10 is viable only for the Dacron material [7, 25, 29]. Therefore, both of these models are compared individually with Modified Lockheed Eq. 4 which is authentic for Silk−net, Dacron as well as Glass−tissue materials [25, 26] at a $P$ value of $10^{-6}$ Torr. It is clear from the left panel of Fig. 3 that for unperforated DAM with Silk−net the heat load varies slowly with hot boundary temperature in Lockheed Martin Flat Plate Eq. 6 in comparison to the Modified Lockheed Eq.4. This is because Lockheed Martin Flat Plate Eq. 6 ignore the gas conduction under vacuum levels that is less than $10^{-5}$ torr [27, 28].

For perforated DAM with Dacron in the case of McIntosh's approach Eq. 10, the value of $\Delta x$ of the spacer material is considered as 5.0 mm for a uniform layout of the radiation shields [35]. The right panel of Fig. 3 exhibits that McIntosh's Layer−by−Layer model has the higher heat load for the perforated DAM with Dacron in comparison to the Modified Lockheed Equation. The shaded region comes due to the range of values for $k$ (0.15 to 0.27 Wm$^{-1}$K$^{-1}$ [30]) which can be seen inside the inset. Thus Modified Lockheed Equation is the better choice for the perforated DAM with Dacron.

Table 1: Empirical coefficients used in the above discussed models for the selected materials. As the Lockheed Martin Flat Plate equation is valid only for Silk−net, that's why constants for Glass−tissue and Dacron are not mentioned.

| Material | Theoretical and empirical Model | $C_r$ $\times 10^{-10}$ | $C_g$ $\times 10^{-4}$ | $C_s$ $\times 10^{-8}$ | $\varepsilon$ | $P$ (in Torr) | Ref. |
|---|---|---|---|---|---|---|---|
| Silk−net | Modified Lockheed | 5.39 | 1.46 | 8.95 | 0.043 | $10^{-4}$ | [25, 26, 32] |
|  | Lockheed Martin Flat Plate | 5.39 | − | 8.95 | 0.03 | $< 10^{-5}$ | [27, 28] |
|  | McIntosh's Layer−by−Layer | − | − | − | $\varepsilon_1 = \varepsilon_2 = 0.04$ | − | [29] |
| 1-7 Glass−tissue | Modified Lockheed | 7.07 | 1.46 | 7.30 | 0.043 | $10^{-4}$ | [25] |
|  | Lockheed Martin Flat Plate | − | − | − | − | − | − |
|  | McIntosh's Layer−by−Layer | − | − | − | $\varepsilon_1 = \varepsilon_2 = 0.04$ | − | [29] |
| 1-7 Dacron | Modified Lockheed | 4.94 | 1.46 | as in Eq.5 | 0.043 | $10^{-4}$ | [25] |
|  | Lockheed Martin Flat Plate | − | − | − | − | − |  |
|  | McIntosh's Layer−by−Layer | − | − | − | $\varepsilon_1 = \varepsilon_2 = 0.04$ | − | [29] |

As the McIntosh's approach Eq. 10 and Lockheed Martin Flat Plate Eq. 6 are effective for certain materials, therefore our current aim of finding a best MLI system with optimal selection of material combination would not go longer with these models. Although all the models discussed above follows their own terms and conditions, Modified Lockheed Eq. 4 provides more degree of freedom in the selection of material criteria because it is authentic for all the material combinations [25, 26]. It follows that the present work considers onwards the Modified Lockheed Eq. 4 to perform the bulk MLI analysis and deliberate the performance of MLI technique as well as it's experimental applicability.

## 3 Results and discussion

Multilayer insulation techniques are often designed in the thermal protection systems to make the heat loads within an acceptable limit. The decrement is analyzed in the value of the heat load, which is coming from the hot wall of the cryostat to the cold wall of the cryostat after implementing the MLI technique [4]. Hence, a good care of MLI technique is required for its better and fruitful performance. To make the MLI technique robust, optimization of various effective parameters are necessary.

The residual gas pressure developed between hot and cold boundaries of the cryostat and emissivity of the radiation shields plays a key role in the improvement of the efficiency of MLI technique. Hence, the effects of the emissivity and residual gas pressure on the thermal performance of MLI technique concerning the heat load has been discussed. In this connection, comparison of heat load between perforated and unperforated radiation shields is performed and achieved optimal perforation style for better performance. As the layering affects the performance of the MLI blenkets [35], therefore the impact of position of the first radiation shield and the arrangement of other radiation shields on the thermal performance of MLI technique has been introduced as useful parameters to evaluate the performance of the MLI technique.

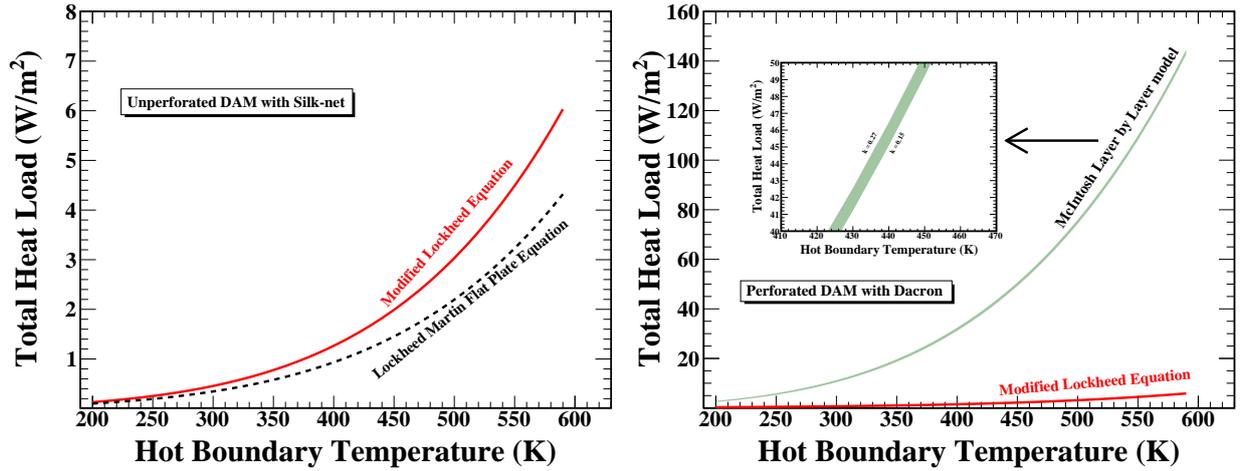

Figure 3: Variation of heat load (in different models) with hot boundary temperature in a cryostat whose cold wall boundary is kept at 77.4 K. Unperforated DAM with Silk−net in the left panel and Perforated DAM with Dacron in the right panel.

## 3.1 Improvement in the performance of the MLI technique

In order to achieve the best thermal performance of thermal insulation i.e. the lowest amount of heat load to the cold wall of the cryostat, the impact testing of the above effective physical parameters is the subject of great interest and importance to thermal and heat transfer technologies. For the investigation of these effective physical parameters, Modified Lockheed Eq. 4 is considered to perform the heat load analysis in bulk as well as the conduction part of the McIntosh's approach from Eq. 10 is considered for the efficiency testing of layering on the conduction part of the heat load. The selected material combinations are defined in the above section. Emissivity is a surface property which strongly depends on material temperature [42], therefore it's analysis with hot boundary temperature id required.

### 3.1.1 Impact of the emissivity of radiation shields

Emissivity is the property of a material which is the ratio of the energy radiated from the surface of a material to that radiated from a black body (a perfect emitter has the value of emissivity = 1). Emissivity depends on the wavelength. According to Planck's law, the total radiated energy increases with the increment in temperature while the peak of the emission spectrum shifts to shorter wavelengths. Thus, it follows that emissivity depends on temperature. As the material gets to a higher temperature, the molecules move more and more, this means they will usually emit more energy [24].

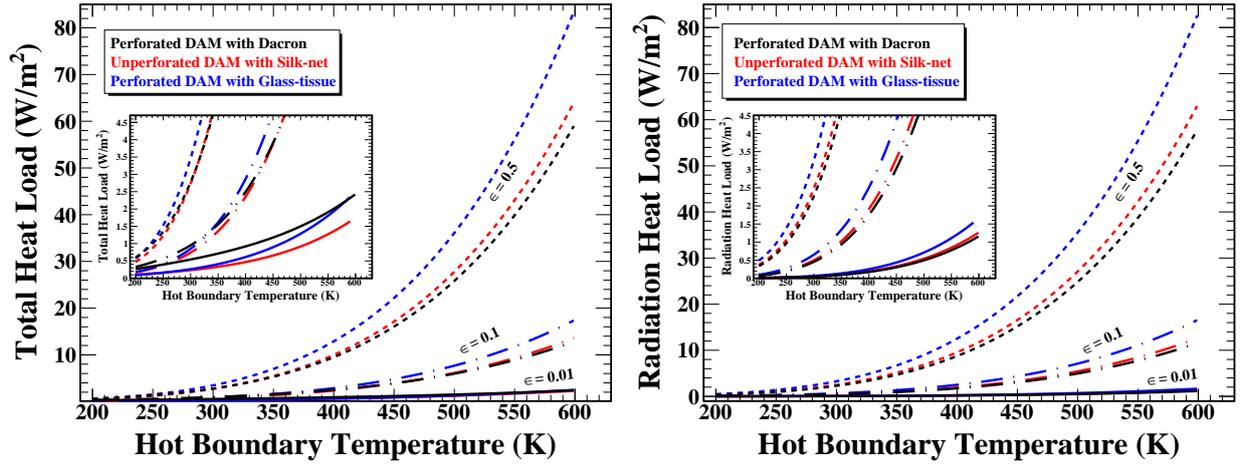

Figure 4: Effect of emissivity of the radiation shields on total heat load in the left panel and the effect of emissivity of the radiation shields on the radiation heat load in the right panel. Both are varying with the Hot Boundary Temperature at the same low boundary temperature of 77.4 K.

Emissivity of the radiation shield in a MLI system is an important parameter which can affect the total heat load (primarily the radiation heat load) on the cold wall boundary of a cryostat. It is directly proportional to the total heat load generated in the MLI technique which is evident from Eq. 4. Generally, Aluminium is preferred for the coating in making radiation shields because it is cheaper than other good conductor materials (Gold, Silver, Copper, etc.) and has quite high melting point 655°C (928 K) [8].

The effect of emissivity on the total heat load as well as radiation heat load can be understood from Fig. 4. To illustrate, the values of $l = 40$, $\bar{l} = 20$ layers/cm and $P = 10^{-4}$ Torr are selected [40, 41], and three experimentally feasible values of $\varepsilon \equiv (0.01, 0.1, 0.5)$ for the selected combination of materials. The selected range for the variation of hot boundary temperature is from 200 K to 600 K and the result is presented in Fig. 4, which represents that the radiation heat load and therefore the total heat load increases (for all the above three selected materials) with emissivity and with the temperature of hot boundary. Mathematically, it is noted from Eq. 4 that the total heat load increases with the increasing value of emissivity because the radiation heat load increases with the increment in the value of emissivity.

Besides this, it happens due to the absorption of maximum incident radiation by the radiation shields, as emissivity increases. This leads to the increment in the temperature of the hot boundary as well as the temperatures of the radiation shields, but a very small amount of heat load reaches to the cold boundary due to multiple reflection of radiation by the radiation shields. This will make the temperature of the cold boundary almost remain constant. Hence, the difference ($T_1$ - $T_2$) increases, which leads to the increment of total heat load as $\varepsilon$ of the radiation shields increases.

It is obvious from Eq. 4 that, $\varepsilon$ is associated with the radiation shields, therefore it's impact on the radiation part of Eq. 4 with warm boundary temperature, is shown in the right panel of Fig. 4. The result conclude that the increment in the radiation heat load as well as total heat load is analyzed maximum for Perforated DAM with Glass−tissue and minimum for Perforated DAM with Dacron. The medium increment is analyzed for Unperforated DAM with Silk−net for all values of $\varepsilon$. This is because of the values of $C_r$ for all the combinations. The highest value of $C_r$ for Perforated DAM with Glass−tissue leads to the greatest increment and the lowest value $C_r$ for Perforated DAM with Dacron leads to the lowest increment in both the heat loads. It is noted from Fig. 4 that there is no major difference in between the total heat load and radiation heat load. This is because, emissivity term appear in the radiation part of the Eq. 4 not in the solid conduction part as well as in the gas conduction part. This leads the greatest impact of $\varepsilon$ on the radiation heat load as compared to the solid conduction and gas conduction heat load.

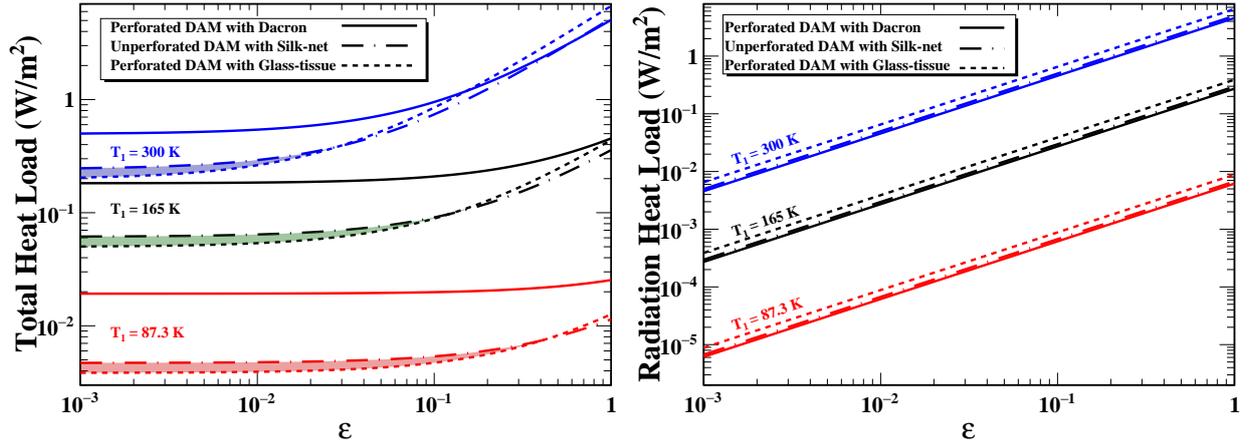

Figure 5: Variation in the total heat load at different hot boundary temperatures in the left panel and the variation in the radiation heat load at different hot boundary temperatures in the right panel. Both the variations are seen with emissivity at 77.4 K of low boundary temperature.

In this sequence, the variation of total and radiation heat load with emissivity is discussed at different hot boundary temperatures and result is shown in the left and right panel of Fig. 5 respectively. This graph specify about the goodness of the selected material combinations in the different range of emissivity. The left panel of Fig. 5 show that, for 300 K of hot boundary temperature, Glass−tissue is best spacer in the emissivity range 0.001 to 0.03, Silk−net is found to be best in the emissivity range 0.03 to 0.6 and dacron is best in the range 0.6 to 1. At 165 K of hot boundary temperature, Glass−tissue is best in the emissivity range 0.001 to 0.15 and Silk−net is found best in the emissivity range 0.15 to 1. At 87.3 K of hot boundary temperature, Glass−tissue is best in the emissivity range 0.001 to 0.4 and Silk−net is found best in the emissivity range 0.4 to 1. It explains that Dacron is worst at the hot boundary temperatures of 165 K and 87.3 K, but it is best at 300 K in the emissivity range 0.6 to 1. Here, it is noticed that the effect of emissivity is significantly estimated as one goes towards the higher values of emissivity.

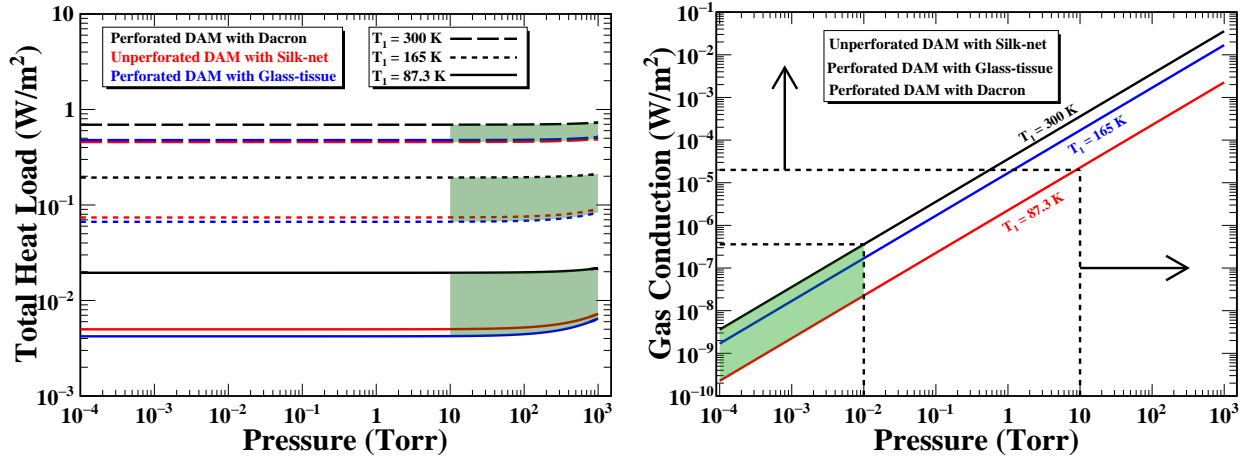

Figure 6: Effect of residual gas pressure on the total heat load in the left panel and the effect of residual gas pressure on gas conduction heat load in the right panel. Both effects are at different hot boundary temperatures with at 77.4 K of low boundary temperature.

### 3.1.2 Effect of the residual gas pressure

Conventionally MLI technique is used under the high vacuum condition level, therefore the heat transfer due to gas conduction is assumed to be negligible under a vacuum level of $10^{-5}$ Torr [27, 28]. However, under the degraded vacuum level, heat transfer due to gas conduction contributes up to a significant amount to the total heat load of MLI technique. It follows that the study of gas conduction heat transfer in MLI technique would play a very important role in improving the efficiency of MLI technique under degraded vacuum.

As it is obvious from Eq. 4 that gaseous conduction is basically influenced by the gas pressure $P$, therefore the variation of the total heat load is analyzed with the gas pressure. To see the effect of gaseous conduction, the range of pressure is chosen from a high vacuum to a poor vacuum which is from $10^{-4}$ to $10^3$ Torr. After incorporating the values mentioned in Table 1, the result is shown in the left panel of Fig. 6, which reveals that at different hot boundary temperatures 300 K, 165 K and 87.3 K, gas conduction start showing it's significance from 10 Torr. Below that pressure value, heat load due to gas conduction is negligible. It means, in the condition of vacuum, gas convection heat load has almost negligible contribution to the total heat load produced and in this situation, the total heat load is only due to the contribution of the solid conduction and the radiation heat load.

To see the effect more precisely, gas conduction part is separately evaluated with $P$ as it is clear from Eq. 4 and presented in the right panel of Fig. 6, which conclude that below $10^{-2}$ Torr, heat load due to gas conduction, is negligible and it starts contributing after 10.0 Torr.

For predicting the impact of the residual gas pressure on the total heat load with hot boundary temperature, three $P$ values $10^{-4}$ Torr, $10^3$ Torr and 1.0 Torr are selected for all the materials. After having all the constant values from Table 1 in the Eq. 4, the result is exhibited in the left panel of Fig. 7, which explains that with the increasing value of $T_1$, the total heat load increases for all the combinations in a similar way. It is also analyzed that the increment in heat load for Perforated DAM with Glass−tissue is found maximum at higher hot boundary temperatures and the increment in heat load for Perforated DAM with Dacron is found maximum at lower hot boundary temperatures. Such behaviour is appearing in the figure by their intersection and is explained by the conductivity of the spacers in low and high−temperature regions [29]. Although all the spacers are compared based on thermal conductivity, hence thermal properties of all these spacers should be clear specifically. Consequently, the conductivity of Dacron is found much greater than for Silk−net and Glass−tissue in the low−temperature region [27], therefore, the total heat load is found maximum for Dacron because of the high conduction heat load at lower hot boundary temperature.

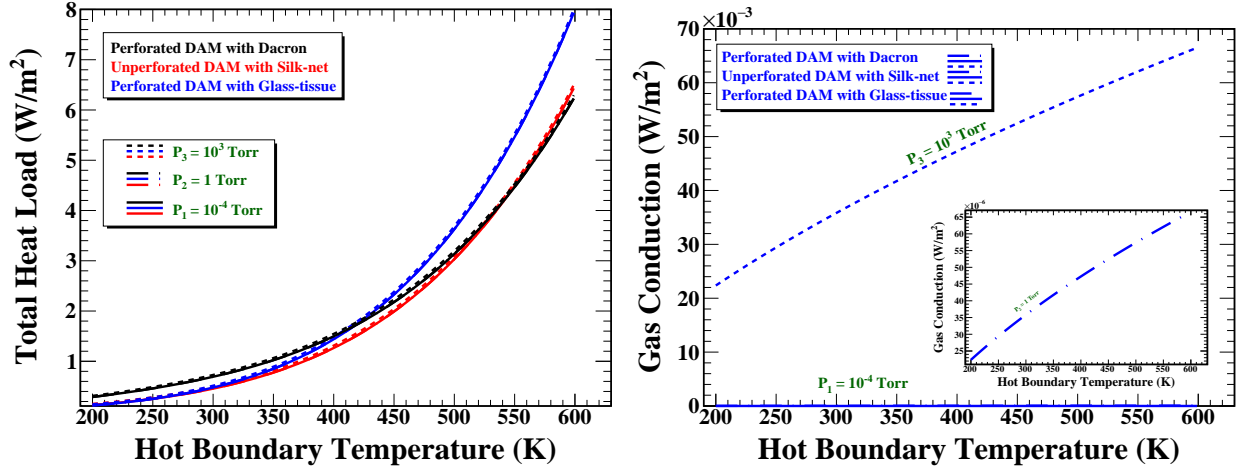

Figure 7: Effect of residual gas pressure on total heat load in the left panel and the effect of residual gas pressure on gas conduction heat load in the right panel. Both these effects are seen with hot boundary temperature and at 77.4 K of low boundary temperature.

It is also seen that $P = 10^3$ Torr has the highest value of total heat load for all the materials because gas conduction is highest at the highest value of gas pressure. As gas conduction is basically influenced by the gas pressure $P$, therefore gas conduction part is separately evaluated with $P$ and presented in the right panel of Fig. 7, which conclude that gas conduction is same for all the three combinations of materials because they all have the same value of $Cg$, but in the left panel of Fig. 7, the total heat load is different for all the three combinations of materials because all the modes of heat transfer are contributing here and all empirical constants are different. Therefore, gas conduction has no significance when the total heat load is estimated and hence, it may be ignored when the total heat load is the point of interest.

### 3.1.3 Significance of perforation and unperforation

Perforation in the thermal shield (radiation shield i.e. DAM) affects the performance of MLI insulation. The heat load is different corresponding to the different fractional open areas in the radiation shields because the perforation of the radiation shield is responsible for heat load as it accommodates a high venting rate which leads to heat load. Accordingly, there are different perforation styles (PS) are possible in the radiation shield as mentioned in the Table 2. The effect of unperforation and perforation of the radiation shield on the heat load is evaluated in the light of Eq. 4 and considering all the constants from Table 2. The performance of MLI technique is analyzed by comparing the heat load produced in Unperforated DAM with Silk−net and the heat load in different styles of perforated DAM with Silk−net. The result is shown in Fig. 8.

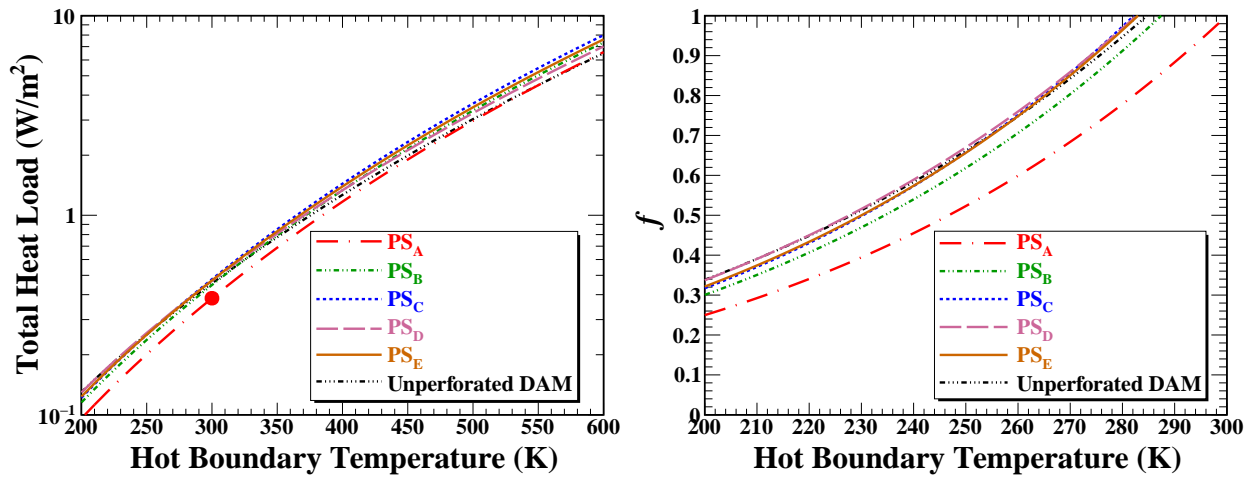

Figure 8: Effect of different perforation styles on the total heat load with hot boundary temperature in the left panel and in the right panel, the relation between the heat load and hot boundary temperature for different perforation styles relative to the heat load of $PS_A$ @ 300 K (water at room temperature).

Table 2: Summary of the key parameters for different perforation styles of radiation shield. All constants are taken from the Ref. [43].

| PS | Hole diameter (cm) | $\varepsilon$ of DAM | Power of $\bar{l}$ in solid conduction term | Fractional open area $\times 10^{-3}$ | Open area (%) | $C_S$ ($\times 10^{-8}$) | $C_R$ ($\times 10^{-10}$) |
|---|---|---|---|---|---|---|---|
| $PS_A$ | 0.119 | 0.043 | 2.84 | 2.60 | 0.26 | 2.98 | 5.86 |
| $PS_B$ | 0.119 | 0.044 | 2.63 | 5.50 | 0.55 | 7.04 | 6.32 |
| $PS_C$ | 0.119 | 0.043 | 2.63 | 10.7 | 1.07 | 7.30 | 7.07 |
| $PS_D$ | 0.229 | 0.042 | 2.35 | 4.80 | 0.48 | 19.9 | 6.10 |
| $PS_E$ | 0.229 | 0.043 | 2.70 | 9.90 | 0.99 | 6.22 | 6.65 |

It is evident from the left panel of Fig. 8 that the different perforated style leads to a higher heat load in comparison to the unperforated DAM, above 550 K temperature. However, below 300 K (the region of interest of MLI technique being used in most of the underground cryogenic experiments) "$PS_A$" provides the least heat load and therefore the best perforation style. Increment in the heat load for different PS is found to vary in same way as their fractional open area increases. This is because the perforation of the radiation shield is accommodating a high venting rate which have a significantly higher heat load > 550 K. Below it, the temperature is not very high and the heat has not been accumulated. Thus, impact of radiation heat transfer is not significant. Therefore the total heat load is not high. It follows that the perforation of radiation shield is responsible for the increment in the heat load in higher temperature region. Optimization of the best perforation style is necessary for being useful in thermal insulation. A relative variation of the heat load with hot boundary temperature for different perforation style with respect to the heat load of $PS_A$ @ 300 K (water at room temperature) is shown in the right panel of Fig. 8. The heat load is quite similar for all the other PS except $PS_A$. Lowest heat load in $PS_A$ leads it to be the best choice if one uses the Perforated DAM with Silk−net as a radiation shield and spacer combination.

### 3.1.4 Arrangement of Radiation shields

It has been analyzed that the layering of radiation shields affect the thermal performance of MLI technique. Layering towards the inner (cold boundary) surface reduces the heat load almost 1.36 and 1.35 times in the case of spherical and cylindrical geometry cryostats, as compared to the layering towards the outer (hot boundary) surface [4]. This is possible because most of the heat incident on the cold boundary from the hot boundary is initially reflected from the radiation shield which is placed near the cold boundary. Remaining heat transmitted to the spacer near the cold boundary, which mostly absorbed in that and not further transmitted towards the cold boundary. This is because of the lower thermal conductivity of the spacer in low temperature region than in the high temperature [29]. Therefore it can not permit most of the heat to reach the cold wall boundary. It follows that the layering of radiation shields in MLI system should be performed by placing the first radiation shield near the cold boundary and successively move towards the hot wall boundary of the cryostat.

However, a non−trivial question arises that how much close or far, the first radiation shield should be placed from the cold boundary to achieve the better functioning of the MLI technique. To address this issue, the conduction heat load in various conditions have been studied because in low temperature−region conduction term dominates the total heat load through the MLI blanket [27]. These different conditions are having the first radiation shield at different positions (or distances) from the side of the cold wall boundary of the cryostat. This is done with the help of the conduction term of Eq. 10. It is because of huge change in conduction due to low thermal conductivity of spacers in low−temperature region [27, 29]. Therefore in the light of Eq. 9, conduction heat load is calculated by decreasing the distance between cold boundary and first radiation shield i.e. decreasing the thickness ($\Delta x$) of the spacer for the first radiation shield.

This analysis is performed for Perforated DAM with Dacron with different cryogenic liquids as Eq. 9 is fitted for Dacron with reasonable accuracy [7, 25, 29]. Therefore the value of $T_2$ for different cryogenic liquids are 77.4 K ($LN_2$) [37–39], 87.3 K for Liquid Argon (LAr) [37, 38] and 165 K for Liquid Xenon (LXe) [38, 39]. The result is presented in the left panel of Fig. 9. It is obvious from the left panel of Fig. 9 that the conduction heat load decreases rapidly with increment in the thickness of the spacer from the cold boundary. This exhibits that placing the first radiation shield at a large possible distance from the cold boundary would provide better result i.e. thickness of the spacer should be large towards the side of the cold boundary. It happens because more heat is transferred from the hot to the cold wall boundary that would be mostly absorbed in the insulation material (spacer) near the cold boundary if the first radiation shield is placed at a large possible distance from the cold boundary. Also, a large thickness of spacer towards the cold boundary region produces a higher thermal resistance [44] which prohibit the transfer of heat (conduction heat load, in particular) coming from the hot boundary. This is because a spacer of high thermal capacity is having low thermal conductivity in the low−temperature region [29, 35]. It is also evident from the left panel of Fig. 9 that the temperature difference $(T_1 − T_2)$ increases by decreasing the temperature of the cold wall boundary and consequently increment in the conduction heat load. This is because a large temperature gradient between the two walls of the cryostat; and the solid conduction is a mode of heat transfer which occurs because of the temperature gradient between the two solid surfaces of a supporting system.

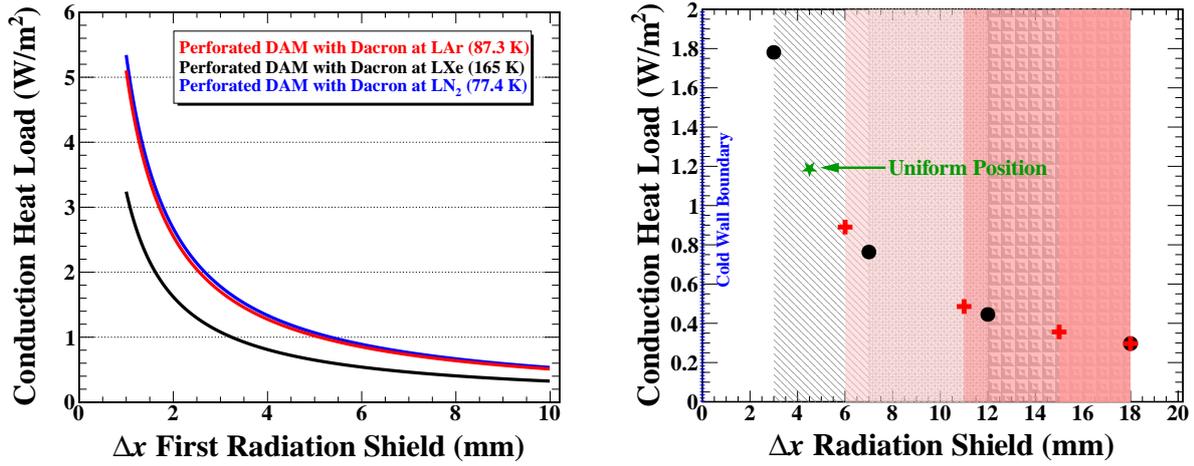

Figure 9: Effect of the position of first radiation shield from the cold boundary in the left panel and the impact of arrangement of radiation shields on the conduction heat load at three cold boundary temperatures in the right panel.

It is clear from the above analysis that the first radiation shield should be placed at a large distance from the cold wall boundary for the better performance of MLI system. However, for the placement of the other radiation shields, it is necessary to arrange them in a suitable configuration to achieve the minimum heat load. This idea comes from the experimental detail that a variable layer density improves the performance of the MLI technique [29]. Calculation for the best arrangement is analyzed with the study of spacing configuration i.e. constant or variable spacing effects of radiation shields on the conduction heat load. In this analysis a uniform spacing i.e. constant layer density, increasing spacing i.e. decreasing layer density and decreasing spacing i.e. increasing layer density from the cold boundary region towards the hot wall boundary region are considered.

The structural layout of the MLI blanket must be analyzed to explore the impact of arrangements on the heat load. In this analysis, an MLI blanket of thickness 20 mm is considered, which contains 0.75 mm sides on both the ends with 5 radiation shields made of DAM sheet of thickness 0.10 mm and 4 spacers with varying thickness according to the above mentioned three different arrangements. A simplified schematic diagram of uniform spacing, increasing spacing and decreasing spacing between the radiation shields is shown in Fig. 10(a, b, c).

1. *Uniform spacing towards the hot boundary region*: Radiation shields are placed with uniform spacing (constant layer density) towards the hot boundary region as shown in Fig. 10(a). In this arrangement, the thickness of the spacer is chosen to be uniform which is 4.5 mm in such a way that: $(2 \times 0.75) + (5 \times 0.1) + (4 \times 4.5) = 20$ mm.

2. *Increasing spacing towards the hot boundary region*: Radiation shields are placed with increasing spacing (or decreasing layer density) towards the hot boundary region as shown in Fig. 10(b). In this arrangement, the thickness of the spacer is chosen in an increasing way such that: $(2 \times 0.75) + (5 \times 0.1) + (3 + 4 + 5 + 6) = 20$ mm.

3. *Decreasing spacing towards the hot boundary region*: Radiation shields are placed with decreasing spacing (or increasing layer density) towards the hot boundary region as shown in Fig. 10(c). In this arrangement, the thickness of spacer is decreasing in such way that: $(2 \times 0.75) + (5 \times 0.1) + (6 + 5 + 4 + 3) = 20$ mm.

All the three arrangements are strongly dependent on the conduction. It follows that the conduction heat load in these three arrangements is studied because it dominates over gas conduction and radiation in the total heat load through the MLI blanket [27] in the low temperature region. These low conductivity spacers are placed between radiation shields (as shown in Fig. 1) to minimize the thermal conduction between the radiation shields through them. This is because the solid conduction became effective and dominant channel over others for the heat exchange [4]. Therefore conduction heat load is analyzed to investigate the performance of the MLI technique with the help of Eq. 9. This analysis is performed for Perforated DAM with Dacron for $T_2 = 77.4$ K (for $LN_2$). The thickness of spacer is varied and consequently position of the radiation shields. To illustrate, in 2nd case, the radiation shields are placed at 3 mm, 7 mm, 12 mm and 18 mm with increasing spacing between them. In 3rd case, they are placed at 6 mm, 11 mm, 15 mm and 18 mm with decreasing spacing between them. Overall, the MLI blanket's space between cold and hot wall boundary remain same of 20 mm. Impact of the uniform, increasing and decreasing spacing and thus the position of radiation shield ($\Delta x$) on the conduction heat load is presented in the right panel of Fig. 9.

It is obvious from the right panel of Fig. 9 that the heat load is constant in the case of a uniform spacing between the radiation shields. A sharp decrement in the conduction heat load is found by increasing as well as decreasing the spacing between the radiation shields. Although the conduction heat load is different for the different position of the radiation shields, it remains same for the whole MLI blanket. A less amount of the conduction heat load is analyzed in the low temperature region for decreasing spacing in comparison to the increasing one. This is because, in the case of decreasing spacing towards the hot boundary region, a large thickness of spacer at cold boundary produces a higher thermal resistance [44]. This high thermal resistance is due to it's high thermal capacity and low thermal conductivity in low temperature region than in the high temperature region [29, 35].

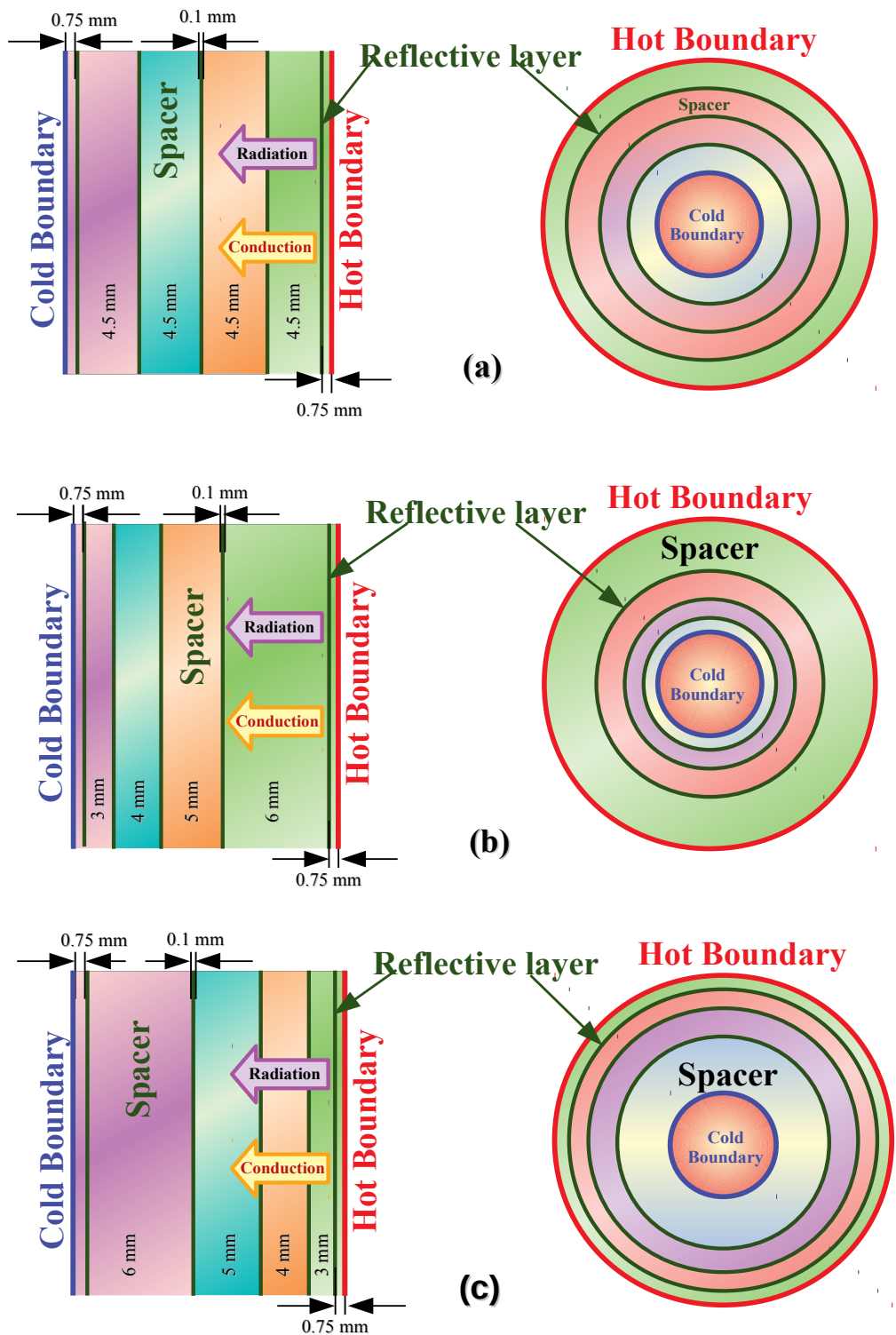

Figure 10: Arrangement of radiation shields: (a) uniform spacing, (b) increasing spacing, and (c) decreasing spacing towards the hot boundary region.

## 3.2 Experimental applicability of MLI technique

Reduction of the heat load is a common challenge in the field of cryogenic applications. A large value of the heat load is undesirable and should be maintained at a minimum possible level. It is quite cumbersome to eliminate the heat load in low−temperature regions. This puts an economic burden and therefore acts as a barrier in cryogenic applications. MLI technique has already been found very useful in reducing the heat load to the cold wall boundary in cryostats [4]. However, an improved and robust MLI technique would play furthermore important role in reducing the heat load. Thus, it turns out to be profitable in preserving the cryogenic liquid by reducing their evaporation rate. Although MLI technique has very vast applications, the present work focuses it's applications in the current running or upcoming projected tonne−scale experiments involved mainly in the search of neutrinoless double beta decay ($0\nu\beta\beta$−decay) [45] such as GERDA, LEGEND−200, LEGEND−1000, CDEX−1T, KamLAND−Zen and nEXO.

### 3.2.1 Contribution in the tonne−scale experiments

The objective of this work is to explore the role of MLI technique in optimizing the consumption rate of the cryo−fluid. It follows that an experiment can run more longer with the same amount of liquid if it's cryostat is equipped with MLI technique. A single layer of radiation shield is inserted near the outer surface as well as near the inner surface to observe the effect of MLI technique. In this scenario, the heat load and the evaporation rate ($E_R$) of cryogenic liquids is evaluated. The temperature of experimentally most applicable cryogenic liquids is listed in Table 3.

Calculation of the heat load requires the shape and size of the cryostats, used cryogenic liquid and the temperature of it's outer surface. The GERDA experiment using a cylindrical cryostat of radius 2.0 $m$ filled with LAr in 64 $m^3$ volume. This cryostat is placed inside a water tank of volume 590 $m^3$ [20, 46]. Same infrastructure of the GERDA would be followed by the LEGEND−200 [47, 48]. The LEGEND−1000 is proposed to use LAr in a cryostat of radius 3.0 $m$ with a volume of 12 $m^3$. This volume would be further divided into four sections each of volume 3 $m^3$ with the help of a thin copper layers [47, 48]. The CDEX experiment is using a cylindrical cryostat of volume 1700 $m^3$ filled with LN$_2$. The height and diameter both of these cryostats are exactly same $\sim$ 13 $m$ [19]. The KamLAND−Zen experiment utilizes a double−walled spherical balloon (filled with LXe) with an inner and outer radius of 1.54 $m$ and 6.5 $m$ respectively [16, 17]. The nEXO experiment is based on the ultra−low background LXe technology and it also has a double−walled spherical LXe cryostat with it's inner and outer radius of 1.69 $m$ and 2.23 $m$ respectively [17, 49].

The heat load for the cylindrical as well as spherical geometry cryostats is calculated following Eq. 1, Eq. 2 and considering input parameters of Table 3. The outcome for different experiments is summarized in Table 4. It is evident from Table 4 that the insertion of a single intermediate layer leads a substantial amount of heat load reduction for all the experiments as well as both geometries. Although cylindrical cryostats are experimentally more feasible, a spherical cryostat offers a less amount of the heat load in comparison to the cylindrical one.

Table 3: Temperature of the cryogenic liquids and corresponding values of their emissivity of mechanically polished copper cryostat.

| Cryogenics Liquids | Temperature (K) | Ref. | $\varepsilon$ | Ref. |
|---|---|---|---|---|
| Liquid Argon (LAr) | 87.3 | [37, 38] | 0.06 | [36] |
| Liquid Nitrogen (LN$_2$) | 77.4 | [37–39] | 0.07 | [11] |
| Liquid Xenon (LXe) | 165 | [38, 39] | 0.3 | |
| Water | 300 | [36] | 0.1 | [11, 36, 39] |

This is due to the multiple reflection of radiation through the MLI technique.

Table 4: Summary of the total heat load in different experiments with and without inserting a single MLI technique's layer for the cylindrical and spherical geometry cryostats.

| Experiment | Isotope | Heat Load (W) | | | | | |
|---|---|---|---|---|---|---|---|
| | | Spherical | | | Cylindrical | | |
| | | without any Layer | layering near R1 | layering near R2 | without any Layer | layering near R1 | layering near R2 |
| GERDA | $^{76}$Ge | 1.43×10$^3$ | 8.45×10$^2$ | 6.21×10$^2$ | 1.67×10$^3$ | 9.81×10$^2$ | 7.27×10$^2$ |
| CDEX−1T | $^{76}$Ge | 1.24×10$^4$ | 7.22×10$^3$ | 5.47×10$^3$ | 1.43×10$^4$ | 8.30×10$^3$ | 6.31×10$^3$ |
| KamLAND−Zen | $^{136}$Xe | 7.24×10$^2$ | 6.82×10$^2$ | 2.53×10$^2$ | 8.25×10$^2$ | 7.45×10$^2$ | 2.92×10$^2$ |
| LEGEND−200 | $^{76}$Ge | 1.43×10$^3$ | 8.45×10$^2$ | 6.21×10$^2$ | 1.67×10$^3$ | 9.81×10$^2$ | 7.27×10$^2$ |
| LEGEND−1000 | $^{76}$Ge | 4.81×10$^2$ | 2.91×10$^2$ | 2.07×10$^2$ | 1.09×10$^3$ | 6.42×10$^2$ | 4.76×10$^2$ |
| nEXO | $^{136}$Xe | 6.86×10$^2$ | 4.56×10$^2$ | 2.78×10$^2$ | 7.70×10$^2$ | 4.93×10$^2$ | 3.19×10$^2$ |

Different experiments are using different volume and thus different surface area of the cryostats. It can be optimize further according to the need of the experiment. Tonne−scale experiments requires a large volume of the cryostat to keep their large amount of cryogenic liquid. A large volume of the cryostat faces more exposure of the heat load over a large surface area in comparison to a small size cryostat. The effect of volume of the cryostat on the heat load is analyzed and the role of a single layer MLI technique. To illustrate, the selected experiments are GERDA and KamLAND−Zen because they are using different cryogenic liquids. This analysis requires to evaluate the surface area corresponding to a given volume of the cryostat. Knowing the surface area, it is done with the help of Eq. 1 (without inserting any layer) and Eq. 2 (with a single layer) using the MLI technique. The result of this analysis is pictorially presented in the left and right panel of Fig. 11.

Insertion of a single layer leads to a significant amount reduction in the heat load. In particular, if this layer is inserted near the inner cold wall boundary $R_2$ then the heat load reduces more in comparison to the near hot wall boundary $R_1$. Enhancement in the volume affects all the heat loads (gaseous conduction, solid conduction, radiation heat load), which is almost similar as well as independent of the geometry of the cryostat. This increment in the heat load with volume occurs because of consequent enlargement in the surface area and it can be understood from Eq. 1 and 2. This variation (without inserting any layer) for all the experiments with their cryogenic liquid and geometry is shown in the below panel of Fig. 11. Quantitatively, this effect is summarized in Table 4 with and without inserting a single layer near the cold and hot wall boundaries. The heat load decreases substantially even after introducing a single layer of radiation shield. Although decrement in the heat load in spherical geometry is more in comparison to the cylindrical one, cylindrical geometry is experimentally more feasible in comparison to the spherical one. It follows that most of the experiments are following cylindrical geometry of the cryostat. Therefore a cylindrical cryostat with even a single layer near $R_2$ would be a good choice for the large−scale experiments.

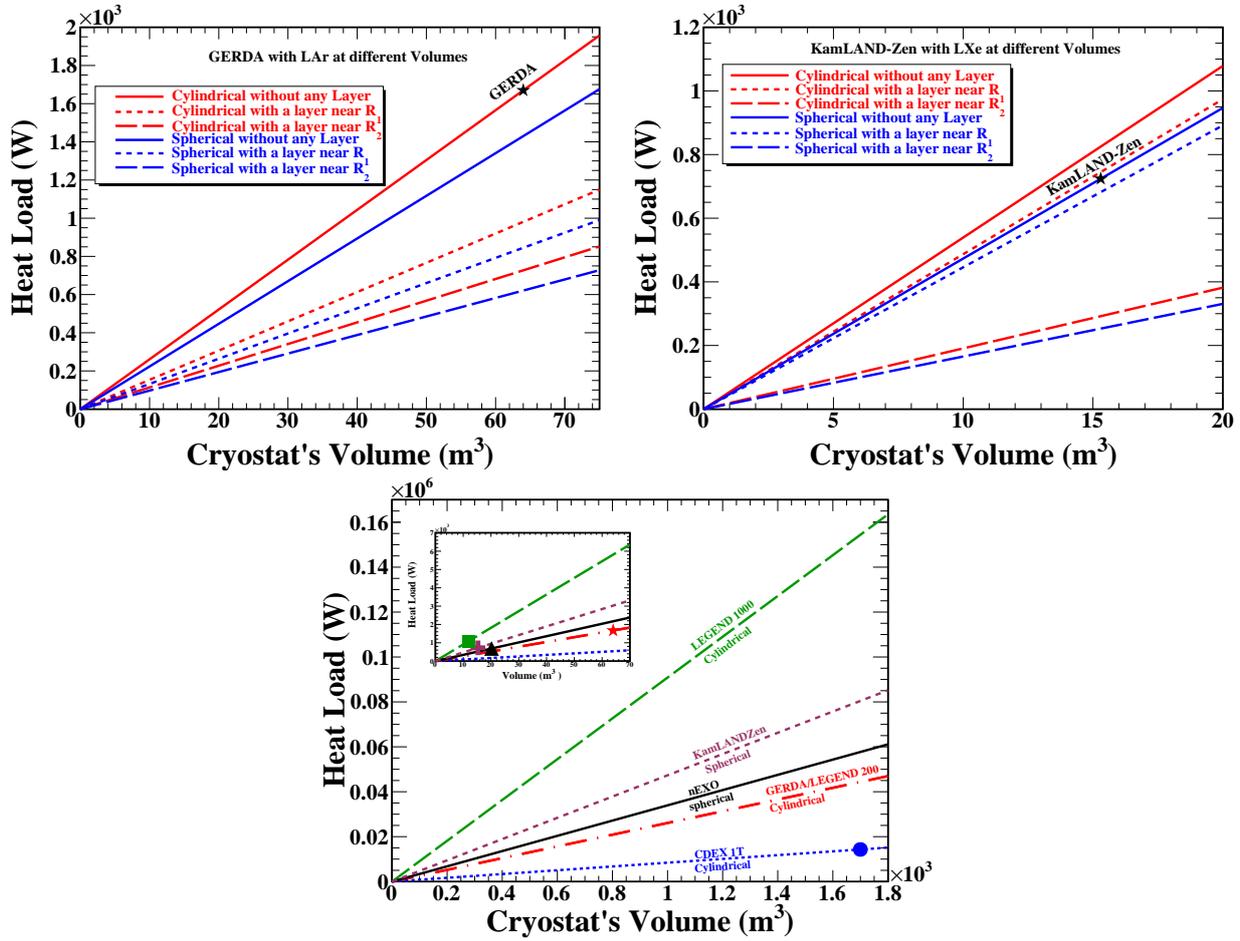

Figure 11: Impact of the cryostat's volume on the heat load: GERDA experiment using LAr cryogenic fluid in the above left panel, KamLAND−Zen experiment using LXe cryogenic fluid in the above right panel and all selected experiments with their current geometry of the cryostat and respective cryogenic liquids without MLI technique in the below panel. Inset in the below panel shows the lower zoomed range of the main figure.

As Silk−net, Glass−tissue and Dacron are used as spacer material with the radiation shields, therefore in the last, the total heat load is evaluated for all the cryogenic liquids with various layer densities in the light of Eq. 4 and is provided below in Table 5. These calculated values shows that after increasing the values of layer densities for all the cryogenic liquids, heat load starts increasing. This happens because the insertion of a large number of radiation shields with insulating spacers results in an increment in the solid conduction heat load between the radiation shields through the spacers, which leads to the increment in the total heat load.

Table 5: Summary table on the calculated total heat load for all liquids at $T_1$ = 300 K and $l$ = 40.

| | Total Heat Load (W/m$^2$)×10$^{-1}$ | | | | | | | | | | | |
|---|---|---|---|---|---|---|---|---|---|---|---|---|
| | For Silk−net | | | | For Glass−tissue | | | | For Dacron | | | |
| $\bar{l}$ = | 10 | 20 | 30 | 40 | 10 | 20 | 30 | 40 | 10 | 20 | 30 | 40 |
| LN$_2$ | 2.53 | 4.56 | 9.17 | 17.13 | 3.13 | 4.78 | 8.54 | 15.03 | 2.76 | 6.93 | 16.38 | 32.69 |
| LAr | 2.52 | 4.51 | 9.03 | 16.83 | 3.12 | 4.74 | 8.43 | 14.79 | 2.73 | 6.72 | 15.79 | 31.44 |
| LO$_2$ | 2.52 | 4.49 | 8.99 | 16.75 | 3.11 | 4.73 | 8.39 | 14.72 | 2.72 | 6.67 | 15.63 | 31.1 |
| LXe | 2.30 | 3.82 | 7.27 | 13.21 | 2.88 | 4.12 | 6.93 | 11.77 | 2.35 | 4.95 | 10.87 | 21.07 |

### 3.2.2 Moderating $E_R$ of cryogenic liquids

The $E_R$ of a cryogenic liquid is directly linked with the heat load. Reduction in the heat load leads to the reduction in the consumption rate of the cryogenic fluid. Evaluation of the $E_R$ of the cryogenic liquid is useful in analyzing the effect of the heat load on the cryogenic liquid concerning it's saving as well as consumption. This analysis follows the evaluation of $E_R$ which requires the information of latent heat of vaporization and density of the cryogenics liquid. These physical parameters and evaluated $E_R$ of cryogenic liquids are listed in Table 6.

Variation in the $E_R$ with the heat load is analyzed and the result is presented in the left panel of Fig. 12. The $E_R$ of LXe is the lowest as compared to the all other listed cryogenic liquids. In other words, LXe is superior than others for the same amount of the heat load. A relative variation of $\left[\frac{E_R}{E_{R_{Xe}}}\right]$ for all liquids vis−a−vis LXe as a reference (corresponding to $E_{R_{Xe}}$ @ 2000 Watt) is shown in the right panel of Fig. 12. This exhibits that LAr and LO$_2$ have quite comparable $E_R$ with LXe thus can be utilized in experiments. Although $E_R$ of LN$_2$ is comparatively larger than LAr and LO$_2$, closer to LXe in comparison to the LNe and LH$_2$.

Table 6: Physical parameters of cryogenic liquids and their evaluated $E_R$.

| Cryogenics Liquid | Latent heat of vaporization (J/g) | Density of Cryogenics Liquids $\times 10^3$ (g/L) | $E_R$ $\times 10^{-2}$ (L/W.hr) | Ref. |
|---|---|---|---|---|
| LAr | 161 | 1.40 | 1.60 | [38, 50] |
| $LN_2$ | 199 | 0.8 | 2.26 | [38, 39] |
| LXe | 95.8 | 2.95 | 1.28 | [39] |
| $LO_2$ | 213 | 1.14 | 1.50 | [38, 51] |
| LNe | 85.8 | 1.21 | 3.50 | [38, 50] |
| $LH_2$ | 445 | 0.07 | 11.5 | [38, 39] |

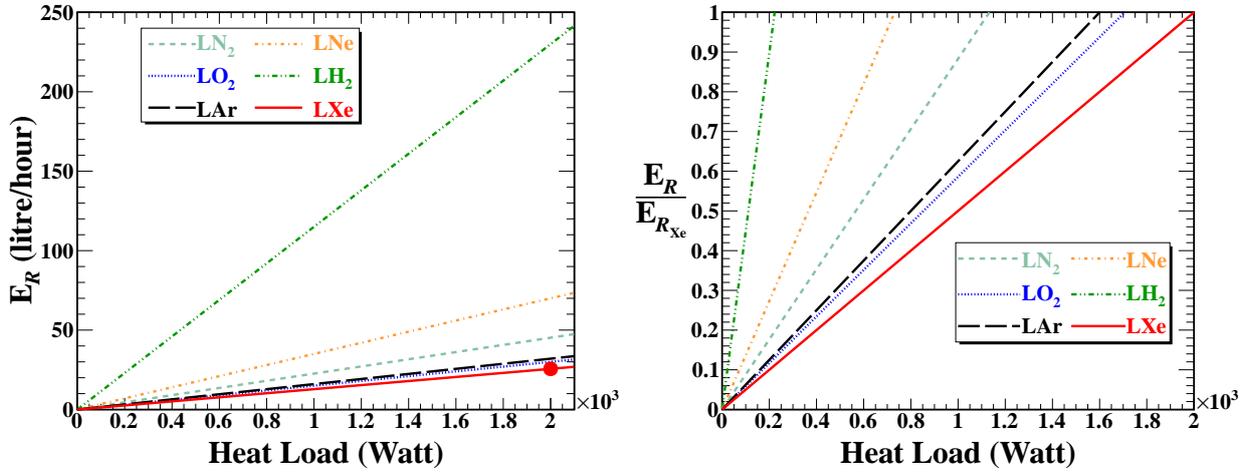

Figure 12: Variation in the $E_R$ of all the selected cryogenic liquids with the heat load in the left panel and a relative variation of the $E_R$ rate with respect to LXe (corresponding to $E_{R_{Xe}}$ @ 2000 Watt) in the right panel.

Enhancement in the heat load leads to the increment in the $E_R$ of cryogenic liquid. From the above analysis, it comes out that the MLI technique is useful in reducing the heat load and therefore in preventing the loss of cryogenic liquid due to boil off. The impact of MLI technique in saving the cryogenic liquid can be understood from Table 7. Saving of cryogenic liquid using MLI technique in five years is summarized in Table 7. Almost > 40 % (single layering near $R_1$) and > 50 % (single layering near $R_2$) cryogenic liquid can be saved with this technique. It is a large amount of cryogenic liquid, which can be saved.

Saving of cryogenic liquid leads to a significant amount of enhancement in the run−time of any experiment. This leads to a substantial amount of reduction in the funding to run a large−scale experiment for long time. Subsequently, a continuous running experiment can accumulate more statistics because of the enhancement in the exposure time. This is what every experiment requires as their one of the most preferred choice. A summary of the experimental run−time enhancement using MLI technique for five tonnes of cryogenic liquid is given in Table 8. Layering near the cold wall boundary helps significantly in reducing the $E_R$

Table 7: Saving of the cryogenic liquid in five years with the help of MLI-technique.

| Experiment | Saving of cryogenic liquid (%) | | | |
|---|---|---|---|---|
| | Spherical | | Cylindrical | |
| | Near R1 | Near R2 | Near R1 | Near R2 |
| GERDA(LAr) | 41 | 57 | 41 | 57 |
| CDEX−1T(LN$_2$) | 42 | 56 | 42 | 56 |
| KamLAND−Zen(LXe) | 5.8 | 65 | 9.7 | 65 |
| LEGEND−200(LAr) | 41 | 57 | 41 | 57 |
| LEGEND−1000(LAr) | 40 | 57 | 41 | 56 |
| nEXO(LXe) | 34 | 60 | 36 | 59 |

and thus increasing the run−time of an experiment.

Table 8: Run time for experiments.

| Experiment | Run time in Days | | | | | |
|---|---|---|---|---|---|---|
| | Spherical | | | Cylindrical | | |
| | No layer | Near R1 | Near R2 | No layer | Near R1 | Near R2 |
| GERDA | 9.5 | 16 | 22 | 8.0 | 14 | 18 |
| CDEX-1T | 0.73 | 1.5 | 1.8 | 0.73 | 1.1 | 1.5 |
| KamLAND-Zen | 23 | 25 | 66 | 20 | 22 | 57 |
| LEGEND-200 | 9.5 | 16 | 22 | 8.0 | 14 | 18 |
| LEGEND-1000 | 28 | 46 | 65 | 12 | 21 | 28 |
| nEXO | 24 | 37 | 60 | 22 | 37 | 52 |

Now comming to fluid loss, it is noted that, to reduce the loss of the cryogenic liquid by heat load, evaporation rates are reduced by using MLI technique. There exists a physical parameter related to cryogenic liquid which can be also changed to find a decrement in the loss of the cryogenic liquid that is volume of the cryogenic liquid. The effect of the volume of the cryogenic liquid on the heat load is thus also important to analyze.

### 3.2.3 Role of MLI technique in saving medical oxygen during current COVID−19 pandemic

The second most abundant (by volume) and prime candidate of the earth atmosphere is Oxygen $O_2$, comprising 20.8 % by volume which is all the way necessary for supporting lives on earth. Though the commercial use of oxygen is primarily in the gaseous form, nevertheless it usually kept in it's liquid form (known as liquid oxygen, $LO_2$) because in doing so, it is economic and to avoid the unnecessary leakage and complications associated with its high−pressure gaseous storage. The vaporizers convert the liquid oxygen into a gaseous state to be used. The Liquid oxygen is a cryogenic liquid i.e. extremely cold with boiling point of 90 K [38, 51, 52] and is very useful in many cases like an strong oxidizing agent, it is used for liquid fuels in the propellant systems of missiles and rockets. Due to having the life−sustaining properties, relied upon in health facilities and medical applications [51, 53].

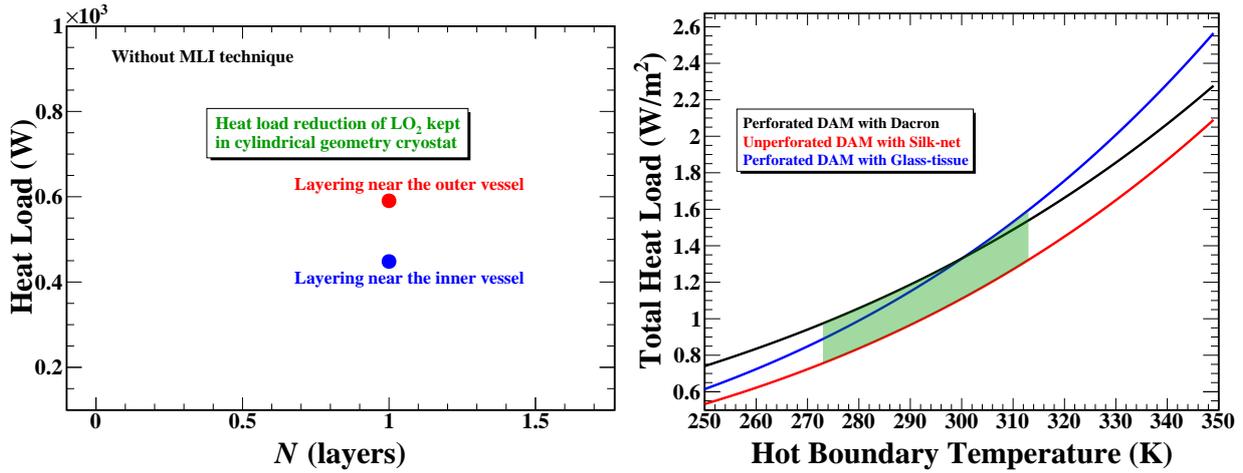

Figure 13: Reduction of heat load with the help of MLI technique in the left panel and the variation of heat load with the storage temperature of LO$_2$ in the right panel.

Oxygen is widely applied in the metal industries in conjunction with acetylene and other fuel gases for metal cutting, welding, scarfing, hardening, cleaning and melting. Steel and iron manufacturers also extensively use oxygen or oxygen−enriched air to affect chemical refining and heating associated with carbon removal and other oxidation reactions. In the chemical and petroleum industries, oxygen is used as a feed component to react with hydrocarbon building blocks to produce chemicals such as alcohols and aldehydes. It is used as a bleaching in pulp and paper industries. It is also used to enhance the combustion rates in glass, aluminum, copper, gold, lead, and cement industries for the waste incineration [51]. Now a days, oxygen therapy is recommended for all severe and critical COVID−19 patients because 15 % of people with COVID−19 are requiring oxygen therapy [54].

After considering its uses, it is extremely essential a special equipment for handling and storage. There is a large temperature difference between the product and the surrounding environment, hence it is essential to keep liquid oxygen insulated from the surrounding heat. A typical storage system consists of a cryogenic storage tanks. These cryogenic tanks are double walled vacuum bottle i.e. refillable cylindrical storage (cryostat) [53], having an inner vessel surrounded by an outer vessel and the space between the two vessels is a vacuum space that contains an insulating medium from which all the air has been removed. This space keeps heat away from the liquid oxygen held in the inner vessel [51].

Since it is impossible to maintain perfect insulation and heat leak is always present there, the inner vessel is trying to draw heat from the outer vessel, vaporization takes place continuously during its use [51, 55] and leads to the loss. If MLI technique is used for the compressed medical oxygen, it will decrease the loss due to heat loads and make it to be more useful for a long time.

The reduction of heat load with MLI technique (inserting a single layer), for the container carries 10,000 liter of liquid oxygen constructed from Stainless Steel [56] is calculated by using

Eq. 1 and Eq. 2 where, the values $\varepsilon_1$ and $\varepsilon_2$ are taken as 0.2 and 0.12 respectively [36]. Since, the ambient temperature for storing $LO_2$ fall in the range between 0-40°C [53], therefore, the values of $T_1$ and $T_2$ is taken as 25°C (298 K) and 90 K [38, 51, 52], having the volume of 10,000 liters with maximum thickness of the container (4 to 8 mm [57]) with height and radius scaled as 4.5 $m$ and 0.84 $m$ respectively, the obtained result is presented in the left panel of Fig. 13.

The left panel of Fig. 13 explain the decrement in heat load by using the MLI technique and the right panel of Fig. 13 gives the best choice for the radiation shield and spacer material. The conclusion obtained from these two plots indicate that for all range of temperature, silk−net is the best choice for spacer material for the storage of $LO_2$. If Dacron and Glass−tissue are the interest of comparison, then the ambient temperature for storage of $LO_2$ is divided into two regions i.e. from 273 K to 300 K Glass−tissue is better than Dacron and above that Dacron is best spacer for MLI technique to save $LO_2$. The second calculated result is that 42 % $LO_2$ can be saved by the layering near the outer vessel and 56 % $LO_2$ can be saved by the layering near the inner vessel of the oxygen tanker in the five years of use.

# 4 Summary and conclusion

High thermal performance of the MLI technique are crucial for reducing the heat load in various applications like storage and transfer operations, cooling of scientific instruments and space launch. In this article, various parameters of MLI systems had been studied which conclude that MLI technique is an advance and typically successful technique which can be used where thermal properties are required to be enhanced. In a cryostat used for low−temperature experiments, the radiation heat load is mainly reduced by placing highly radiation shields, separated by low thermal conductivity materials. This analysis has been performed with the selected material combinations of radiation shields and spacers.

The performance of the MLI technique is greatly influenced by the emissivity of the radiation shields and hence, emissivity effect has been analyzed on the thermal performance of MLI technique by using the Eq. 4. It comes out that, with the increasing value of emissivity, radiation heat load increases and hence the total heat load. It is concluded that the low emissivity of radiation shields is more precise to use to get better performance. In this study, the effect of residual gas pressure on the thermal performance of the MLI technique has been analyzed and the result reveals that the lowest pressure value (intense vacuum) would provide better insulation because of the ignorance of gas conduction heat load. The effect of perforation is also an important ingredient for the performance testing purpose and concluded that perforation of radiation shield would provide the higher value of heat load as compared to the unperforated radiation shields, above the temperature range 550 K. The best perforation style is "$PS_A$" in the region of interest of cryogenics (temperature < 300 K) used in the ground experiments, which is evaluated on the basis of the heat load comparison between all the perforation styles.

The thermal insulation performance of the MLI technique is greatly changed with the effect of the position of the first radiation shield, from the cold boundary of the cryostat. Eq. 9 conclude that the thickness of the spacer should be large enough towards the cold boundary region, to get the better performance. In this sequence, after fixing the position of the first radiation shield, it can be concluded that the other radiation shields should be arranged in decreasing the spacing pattern of radiation shields towards the hot boundary region. This arrangement is best because of the large thickness of spacer towards cold boundary region. The conduction heat load is got prevented to reach to the cold wall boundary of a cryostat, due to having the low thermal conductivity of spacer in the low−temperature region.

The applicability of the MLI technique is evaluated, in the reduction of heat load after inserting a single layer of the radiation shield for both the shapes of the cryostat. A small amount of heat load is offered by the spherical shape as compared to the cylindrical shape of the cryostat. After calculating the evaporation rate of all cryogenic liquids, it was noticed that the best cryogenic liquid is LXe with lowest evaporation rate and can be used for a long for the temperature control. The % of saving of cryogenic liquid is shown in Table 7 which explains that a large amount of cryogenic will be saved in five years with the help of MLI technique. It leads to an increment in the run time of an experiment which is shown in Table 8.

To reduce the loss of the cryogenic liquid, the heat load is reduced by using the MLI technique. Besides this, the volume of the cryogenic liquid can be controlled to reduce it's loss. The effect of volume of the cryogenic liquids on the heat load, used in any experiment, is evaluated which concludes that a small amount of cryogenic liquid should be used to get a low amount of heat load.

Finally, it is evaluated that for all the range of ambient temperature of storage for medical oxygen, Silk−net is the best choice for spacer material. The saving of LO$_2$ for five years of use, is calculated, which is 42 % with the layering near the outer vessel and 56 % with the layering near the inner vessel of the oxygen tanker.

# 5  Outlook

Having identified a credible route to do a reliable investigation one often encounters situations where experimental wisdom from one paradigm fails to match the same in other paradigm. In areas of cryogenics this may turn out to be the case as one performs activities at extreme temperature, either very low or very high. Or moves from microscopic to macroscopic domains.

Theoretical estimates that often serves as good guidelines for any experiment, my go however depending on the complexity of a situation. We try to point out a possible way to handle such a situation here.

The Microscopic transport coefficients (MTC) and their constituents like electrons, phonons etc. show a wide range of variation in their properties with temperature, or pressure or other intensive or extensive thermodynamic variables. In spite of this, it has been observed that, the ratio of certain quantities show a universal behaviour when temperature is taken to zero. One of the prominent example to cite this Widemann Franz's law [58]. Apart from this there is a law that claims universal ration of heat conductivity to viscosity at moderate and low temperatures. And lastly the famous universal ratio of viscosity $\eta$ to entropy $s$ that had been claimed to reach $1/4\pi$ in the famous references [59–61]

The relation of Widemann Franz's law states,

$$\frac{K}{\sigma} = \frac{8}{\pi}\frac{T}{e^2} \qquad (11)$$

where, $\sigma$ happens to be the electrical conductivity. The relation of [62] states,

$$\frac{K}{\eta} = \frac{6}{M} \tag{12}$$

where M stands for the molecular weight of the constituent. Both these relations are obtained using condensed matter models.

On the other hand, recalling that at constant volume entropy, $s \sim \rho$, where $\rho$ is the internal energy. We can have a rough estimate of $\eta/s$ by expressing the viscosity $\eta$ in terms of the mean velocity $\bar{v}$ and mean collision time $\tau$ as $\eta \sim m\bar{v}^2\rho\tau$. Upon evaluating the ratio $\eta/s$ in this approximation we find the disappearance of the energy density $\rho$ and we are left with, $\eta/s \sim m\bar{v}^2\tau \sim E\tau \sim 1/2$ (heuristically). Where in the last line we have used the energy time uncertainty relation and finally considered $\hbar$ to be equal to unity, which is the case in natural units. The viscosity to entropy ratio evaluated in [59], was found to be one over four pi. Therefore we can consider,

$$\frac{\eta}{s} = \frac{1}{4\pi} \tag{13}$$

If these relations are true one can express each of the coefficients in terms of the entropy. Since according to thermodynamics, entropy for all processes must be greater or equal to zero, then these relations should ensure that the respective coefficients should also remain positive or zero, whether estimated in laboratory or theoretically. Though this program has not been implemented so far to the best of our knowledge, as of now; but it holds the promise of being a useful program for future.

## Acknowledgements

The author D. Singh sincerely acknowledges Council of Scientific and Industrial Research (CSIR−UGC), New Delhi, India, for the financial support in the form of CSIR (JRF/SRF) fellowship. The authors D. Singh and V. Singh are grateful to the Ministry of Human Resource Development (MHRD), New Delhi, India for the financial support through Scheme for Promotion of Academic and Research Collaboration (SPARC) project No. SPARC/2018−2019/P242/SL.

# 6 Appendix

# A Initiation to Heat Transmission

The type of heat transport we encounter around us are basically of three types. To name them individually they are as follows, (1) radiative transport, (2) conductive transport and (3) Convective transport. Usually radiative and conductive transport doesn't involve physical transport of material (except for the charge carriers they are composed of) however convective transport do involve material transport. The transportation of the third kind, actually involves effects due to viscosity during transportation on grounds of fluid like nature of the materials that transport.

## A.1 The different ways understand transportation

The most general microscopic set of equations those are capable of describing the ordinary transport processes taking place around us, at normal temperature and pressure, are named after Ludwig Boltzmann, to honour his contribution in this area of science. This approach of Boltzmann doesn't consider the materials under consideration to be in equilibrium. So in some sense the systems those are far away from equilibrium can be described in terms of them. On the other hand assuming the system to have reached local thermal equilibrium, a set of equations, called fluid or hydrodynamic equations (called by the name Navier Stokes or Euler Equations) are also used at times to describe energy transport. As and when the non−equilibrium effects are considered, these equations provides results those can match real life experiments. In this note, in stead of considering these equations in their full generality, we would restrict ourselves to those parts that can be considered to describe transportation in small scale systems. That is to say the systems where the size of the perturbation on the system and its response thereafter are negligible to the size of the system.

# B Types of heat transport

Usually heat or energy transport are of three types, rdiative transport, conductive transport and convective transport. Out of the three radiative heat transport doesn't require presence of any medium. At high temperature it moves in straight line thus the radiated area is usually bears a similarity to the shape and size of the source. At low temperature particularly in the far infrared region, the wavelength of the light ray becomes large. Hence even if there are obstacles in its passage it propagates following the principles of diffraction.

Conduction on the other hand takes place with the help of the free electrons and phonons. The relevant parameter happen to be the relaxation time. Lastly, convection takes place following the principles of fluid motion be it stream−lined or turbulent. The notable point of fluid motion at low temperature is, even at low temperature, depending on density the same can be dissipative and the dissipation can take place through vortex formation. The rate of heat transfer in this process depends linearly on the difference between the temperature of the boundary wall and the fluid in question.

The energy transfer process, usually takes place following a smooth laminar structure unless the perturbations to the system are negligible. The perturbations can be mechanical or thermal like temperature gradient $\frac{dT}{dx}$ etc. For cryogenic applications, the temperature gradients are usually pretty high hence some measures like MLI, as discussed in the text, needs to be taken.

The heat transmission in this process also follows the three process discussed above. The amount of heat transmission however, varies from process to process. In the following we shall revived them and try to suggest some consistency checks that might be useful to check the performance of the device.

# C  Radiative transmission

Radiative heat transfer is a surface based phenomena, that depends on the characteristics of the surface of body. In this phenomena usually three processes (i) absorbtion, (ii) emission and (iii) reflection, take place. The processes of emission and absorption involves the internal energy $U$. Following the principles of thermodynamics, this quantity turns out to be the equal to each other. The parameter called emissivity is usually denoted by $\varepsilon$ and is relates the temperature $T$ and the energy $E$ and the Stefan−Boltzmann's constant $\sigma$, by the following relation,

$$E_i = \sigma \; \varepsilon \; T^4 \tag{14}$$

Numerically, the magnitude of $\varepsilon$ ranges between 0 to 1. As mentioned above the absorption coefficient should also be the same as emission coefficient. The subscript $i$ in Eq. 14 stands to identify the emission power of a location denoted by "i".

We will use physics of emission absorbtion and reflection to find the heat transport taking place between two mutually facing foils of a MLI. Thus if certain amount of energy $E_L$ falls on the left surface of an MLI foil having emissivity $\varepsilon_L$, then $\varepsilon_L E_L$ is the amount of energy that will be absorbed by the concerned foil and $(1 - \varepsilon_L)E_L$ amount of energy will be reflected back towards the right foil. We assumed the foils to have unit areea and they are parallel to each other.

So the total emissive power (emission plus reflection) for the left foil if denoted by $T_{eL}$, the same would be given by,

$$T_{eL} = \varepsilon_L E_L + (1 - \varepsilon_L) T_{eR} \tag{15}$$

Similarly, the total emissive power of the surface on the right side given by $T_{eR}$ and would turn out to be,

$$T_{eR} = \varepsilon_R E_R + (1 - \varepsilon_R) T_{eL} \tag{16}$$

If we solve the two Eqs. 15 and 16, we would get the expressions for $T_{eL}$ and $T_{eR}$. If we take the diference or the two, we would arrive at the expression of the net heat transfer which is given by

$$q_{L-R} = \frac{E_L - E_R}{\left(\frac{1}{\varepsilon_L} + \frac{1}{\varepsilon_R} - 1\right)} \tag{17}$$

For the special situation when the material for left and right foil happen to be the same then we should have,

$$q_{L-R} = \frac{E_L - E_R}{\left(\frac{2}{\varepsilon_L} - 1\right)} \tag{18}$$

if we use $N$ number of foils to reduce the heat trnsmision (load) the factor of two in Eq. 19 would instead be replaced by a factor of $N$ and expressed as

$$q_{L-R} = \frac{E_L - E_R}{\left(\frac{N}{\varepsilon_L} - 1\right)} \tag{19}$$

In the limit of $N$ tending to infinity, one can see that the final heat load would turn out to be zero. That means one has perfect insulation. However, there is a caveat in the argument, that is we have not kept track of the geometry of the device in the limit N taken to infinity. As it is easy to guess in this limit the size of the device would become huge. Thus we would loose all the operational benefits. It is therefore necessary to do some optimization, that is done by simulation.

## D  Transport coefficients and the constraints

### D.1  conduction and convection

Similarly, one can try to estimate the convective and the conductive heat transport for MLI. Under some simplified assumptions, the heat load for conductive and convective transport has been found to be[4],

$$q_{\text{condt+convect}} = \frac{C_S \ \bar{N}^{2.63} \ (T_1^2 - T_2^2)}{2 \ (N+1)} + \frac{C_G \ P \ (T_1^{0.52} - T_2^{0.52})}{N} \qquad (20)$$

where, $\bar{N}$ stands for layer density, $N$ is the number of layers and the constants $C_S$ and $C_G$ stands for the coefficient of solid conduction and gas conduction, respectively.

Transport of energy or matter in a medium depends on the transport agents those transfer the items. A fluid equation bassed approach, using euler or Navier stokes equation, to find the energy transport through conductive and convective processes, had been available in the literature for some time now. A rigorous introduction to the same in the astrophysical context, can be found in [63] and the references there in.

Using this formalism with all the details is a difficult job. However some simplifications are possible under some simplifying assumptions. Approximate expression obtained under these simplifying assumptions were found and are [7]. The same were used further in [4] for the study of heat load in cryogenic devices.

It was found there that these macroscopic heat transport equations depend on heat transport coefficients. These coefficients in turn depend on the properties of the carriers that carry the heat energy or any other form of energy in a material. For solid material they may be electrons, phonons etc. and in fluids they may be composed of the materials that make the fluid up.

The transport coefficients can thus depend on the viscosity ($\eta$) or the thermal conductivity ($K$). These fluid equations finally should respect the thermodynamic constraints that the thermodynamic variables are supposed to. Like for any natural process entropy must be greater than or equal to zero. Therefore the coefficients $C_S$ and $C_G$, in eqs. (3.2) (3.6) of [4], as well as Eq. 20 of the current paper, should depend on $K$ and $\eta$. So to model the performance of the thermostats for cryogenic materials it would be a good initiative to start using the exact transport coefficients in the transport equations and verify the matching of experimental observation with the theoretical prediction. It may turnout to be a good way maintain the credibility of ones work.